\documentclass[twocolumn,showpacs,amsmath,amssymb, aps, prl, showkeys]{revtex4-1}

\usepackage{graphicx}
\usepackage{natbib}

\begin{document}


\title{High-order optical nonlinearity at low light levels}


\author{Joel A. Greenberg}
\email[]{jag27@phy.duke.edu}
\affiliation{Department of Physics and the Fitzpatrick Institute for Photonics, Duke University, Durham, NC 27708, USA}
\author{Daniel J. Gauthier}
\affiliation{Department of Physics and the Fitzpatrick Institute for Photonics, Duke University, Durham, NC 27708, USA}


\date{\today}

\begin{abstract}
We observe a nonlinear optical process in a gas of cold atoms that simultaneously displays the largest reported fifth-order nonlinear susceptibility  $\chi^{(5)}=1.9\times10^{-12}$ (m/V)$^4$ and high transparency.  The nonlinearity results from the simultaneous cooling and crystallization of the gas, and gives rise to efficient Bragg scattering in the form of six-wave-mixing at low-light-levels.  For large atom-photon coupling strengths, the back-action of the scattered fields influences the light-matter dynamics.  This system may have important applications in many-body physics, quantum information processing, and multidimensional soliton formation.
  
\end{abstract}


\maketitle

Since the first observation of second-harmonic generation of a ruby laser beam over 50 years ago, there has been  sustained effort to improve the nonlinear optical (NLO) interaction strength of materials. One ultimate goal is to realize nonlinear interactions at the single-photon level, which will lower the operating power of devices. There is also a growing need for single-photon nonlinearities for quantum information applications \cite{kimble08}.

The strength of a material's nonlinear optical response is characterized by the $n$-th order susceptibility tensor $\overleftrightarrow{\chi}^{(n)}$, which relates the nonlinear polarization of the material $\vec{P}^{NL}$ to the electric field strength $\vec{E}$  via $\vec{P}^{NL}=\epsilon_0 [\overleftrightarrow{\chi}^{(3)}:\vec{E}\vec{E}\vec{E}+\overleftrightarrow{\chi}^{(5)}:\vec{E}\vec{E}\vec{E}\vec{E}\vec{E}+...]$ for an isotropic material \cite{boydnlo}. Because the magnitude of the susceptibility typically decreases with increasing order, most low-light-level studies to date have focused on lower-order processes.  For example, there have been numerous observations of low-light-level NLO interactions based on third-order ($\chi^{(3)}$) processes created via electromagnetically-induced transparency (EIT) \cite{hau99, shiau11, sevincli11, lo11}, where a strong coupling beam creates a quantum interference effect that simultaneously renders the medium nearly transparent while enhancing the nonlinearity \cite{fleischhauer05}.  Other recent observations of strong third-order NLO effects include a two-photon absorptive switch actuated using $<20$ photons interacting with atoms in a hollow-core photonic bandgap fiber \cite{venkataraman11} and an optical pattern-based switch using $\sim$600 photons interacting with a warm atomic vapor \cite{dawes10}. 

Despite the success of these approaches, some applications require or can benefit from higher-order nonlinearities.  Materials with a large,  fifth-order ($\chi^{(5)}$) response, for example, can lead to new phenomena, such as liquid light condensates \cite{michinel06} and transverse pattern and soliton formation \cite{fibich07}, and play an important role in quantum information networks through enabling 3-qubit quantum processing \cite{hang06,zubairy02}, providing new sources of correlated pulse pairs \cite{felinto10}, and acting as quantum memories \cite{felinto10,kang04}.  Quintic media can also improve high-precision measurements \cite{zhang09} and reduce phase noise for enhanced interferometry performance \cite{giri03}.  Thus, the realization of efficient $\chi^{(5)}$ materials is important for fundamental studies in quantum nonlinear optics as well as improving the performance of NLO devices.   

In this Letter, we report the discovery of a \textit{dissipation-enhanced} NLO process that gives rise to the largest fifth-order ($\chi^{(5)})$ NLO susceptibility ever reported while simultaneously having high transparency.  Our NLO material consists of a gas of cold atoms initially in thermal equilibrium, which is illuminated by weak, frequency-degenerate laser beams (frequency $\omega$).  The light fields (both applied and self-generated via wave mixing) act on the atomic center-of-mass motion to cool and crystallize the gas and lead to Bragg scattering via the generated atomic density gratings.  This cooling arises via the dissipative Sisyphus force \cite{dalibard89} and occurs efficiently (\textit{i.e.}, requires the scattering of only tens of photons per atom) so that absorption can be made small.   Thus,  in contrast to previous studies of wave mixing via atomic bunching that produce a third-order NLO response \cite{inouye99,kruse03,  schilke11}, dissipative effects are crucial to the nonlinearity discussed here and cause the lowest-order nonlinearity to be fifth-order in the applied fields.

Surprisingly, the achievable nonlinear interaction strength observed in our experiments from the $\chi^{(5)}$ response is just as large as that obtained in previous experiments dominated by a $\chi^{(3)}$ response.  This strong light-matter coupling enables the scattered fields to alter the  atomic spatial organization and momentum distribution, resulting in greatly reduced group velocities, enhanced atomic coherence times, and optical instabilities.  Because of this back-action, our system may be particularly interesting for studies of strongly-correlated many-body physics with long-range interactions \cite{gopal10}.

\begin{figure}[h, t]
 \includegraphics[width=3.in]{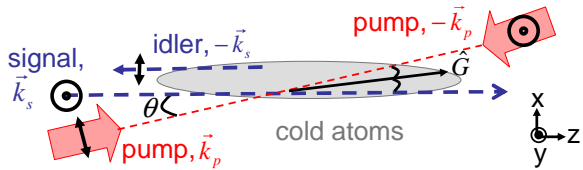}
 \caption{Schematic of the experimental setup.  We take $z=0$ ($z=L$) to be the left (right) side of the cloud.  }
 \label{fig:setup}
 \end{figure}

To make our discussion  concrete, we consider the situation shown in Fig. \ref{fig:setup}, where optical fields interact with a cloud of cold atoms in a pencil-shaped geometry.   A pair of strong, counterpropagating pump fields (intensity $I_p$, wave vectors $\pm \vec{k}_p$) are inclined by an angle $\theta$=10$^{\circ}$ relative to the cloud's long axis and are linearly-polarized with orthogonal polarizations (lin$\perp$lin configuration).  A weak signal field (intensity $I_s$, wavevector $\vec{k}_s$) is injected along the $z$-axis with a polarization parallel to the nearly-counterpropagating pump field.  An idler field ($I_i$, -$\vec{k}_s$), self-generated via NLO wave mixing, counterpropagates with and has a polarization that is orthogonal to the signal field. We refer to the signal and idler fields as probe fields.  

In our experiment, we use an anisotropic magneto-optical trap (MOT) to confine $^{87}$Rb atoms in the $5 \hspace{3 pt}^2S_{1/2}(F=2)$ state within a cylindrical region of length $L$=3 cm (along $\hat{z}$) and diameter $W$=300 $\mu$m (along $\hat{x}$ and $\hat{y}$)  \cite{greenberg07}.  This configuration enables us to achieve typical atomic densities of $\sim5\times10^{10}$ cm$^{-3}$ for atoms isotropically cooled to $T_{eq}$=20-30 $\mu$K. The pump and probe beams are tuned below the $5 \hspace{3 pt}^2S_{1/2}(F=2)\rightarrow 5 \hspace{3 pt}^2P_{3/2}(F'=2)$ transition (transition frequency $\omega_{23}$) by $|\Delta|/\Gamma=3-20$ ($\Delta=\omega-\omega_{23}$) and have diameters of 3 mm and 200 $\mu$m, respectively. We use pump beam intensities of up to a few mW/cm$^2$ and a signal beam intensity of 3 $\mu$W/cm$^{2}$.  Thus, all beams are well below the off-resonance electronic saturation intensity $I^{\Delta}_{sat}=I_{sat}[1+(2\Delta/\Gamma)^2]$, where $I_{sat}=4 \epsilon_0 c \hbar^2/(\mu^2\Gamma^2)=1.6$ mW/cm$^2$ is the resonant saturation intensity, $\Gamma/2\pi=6$ MHz is the natural transition linewidth, $\mu=2.53\times10^{-29}$ C$\cdot$m is the reduced transition dipole moment, $\epsilon_0$ is the permittivity of free space, and $c$ is the speed of light in vacuum. 

To measure the system's NLO response, we cycle between a wave-mixing and a cooling and trapping phase.  During the wave-mixing phase, we turn on only the pump and signal beams for $\sim1$ ms and record the time-dependent  intensities of the signal [$I_s(z=L)$] and idler [$I_i(z=0)$] beams as they exit the cloud.  After $\sim200$ $\mu$s, the system reaches a steady state that persists for $\sim1$ ms (at which time expansion of the cloud in the $y$-direction reduces the density and, consequently, the nonlinearity).  We then cool and trap the atoms with only the MOT beams for 99 ms and repeat the cycle.  We have verified that the MOT magnetic fields, which remain on during the experiment, do not affect the NLO response \cite{greenberg11}.  

\begin{figure}[b, h, t]
\centering
 \includegraphics[width=3.4in]{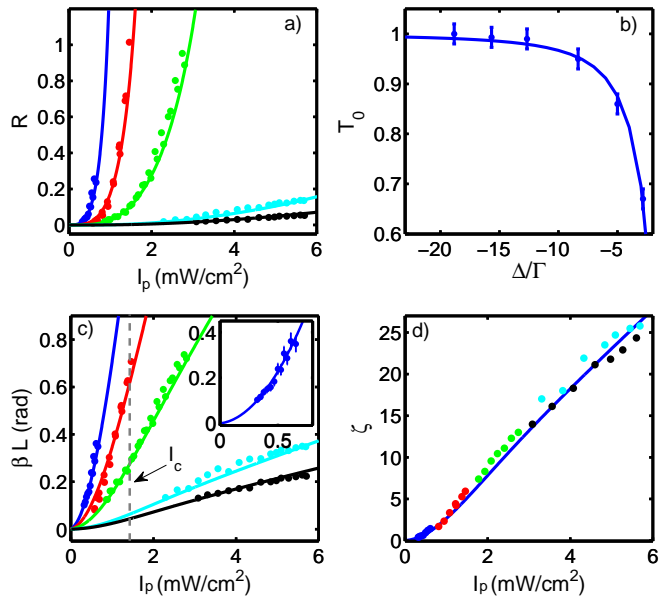}
 \caption{a) Dependence of $R$ on $I_p$ for $\Delta/\Gamma=$ -3, -5, -7.3, -12.7, and -18.8 (from left to right, respectively).  b)  Signal-beam transmission for $I_p=0$.  c) Nonlinear phase shift corresponding to the data in a) obtained via Eq. 4.  The vertical dashed line indicates the value of the characteristic intensity $I_c$.  The inset shows that, for $I_p \ll I_c$, $\beta$ is well-fit ($r^2$=0.994) by a quadratic function (where we show results for $\Delta/\Gamma=-3$).  d)  Cross-phase modulation figure of merit as a function of $I_p$ for the same detunings as in a).  The fact that all points follow the same curve indicates that both $\beta$ and $\alpha_\Delta$ scale as $(\Delta/\Gamma)^{-2}$.  For all figures, $\alpha_{\Delta=0} L=13.4 ( \pm 0.5)$ (uncertainty corresponds to one standard deviation), the points correspond to the experimental data, and the solid curves correspond to the theory via Eqs. \ref{eq:beta} and \ref{eq:soltns}.}
 \label{fig:scaling}
 \end{figure}

Figure \ref{fig:scaling}a shows the steady-state reflectivity $R=I_i(0)/I_s(0)$ as a function of $I_p$ for different $\Delta$, where the points (solid curves) correspond to experimental measurements (theoretical predictions, which we discuss later). The reflectivity scales  superlinearly with $I_p$ and can approach 1 for weak pump intensities (\textit{i.e.}, $I_p \ll I_{sat}$).  Beyond $R\sim 1$, an optical instability occurs that gives rise to self-generated signal and idler fields in the absence of an injected signal beam \cite{greenberg11}.  This instability, coupled with the rapid variation of $R$ on $I_p$, ultimately limits the largest values of $R$ that we can measure. 

For a fixed $I_p$,  $R$ decreases with increasing $|\Delta|$.  Nevertheless, this decrease occurs sufficiently slowly that we can still obtain large $R$ while operating far from resonance (where absorption is minimal).   Figure \ref{fig:scaling}b shows  the signal-beam transmission $T=I_s(L)/I_s(0)$ in the absence of the pump beams, given by $T_0=\text{exp}(-\alpha_\Delta L)$, where $\alpha_\Delta=\eta \hbar \omega \Gamma/2 I^\Delta_{sat}$ is the off-resonance absorption coefficient and $\eta$ is the average atomic density.  For all of the detunings shown in Fig. \ref{fig:scaling}a, $T_0>0.65$ and approaches $1$ for the larger detunings.  Thus, this light-matter interaction allows us to realize large nonlinearities with high transparency for low input intensities (\textit{e.g.}, we measure $R=1$ for $I_p=1.5$ mW/cm$^2$ by working at $\Delta=-5 \Gamma$ where  $T_0=0.87$).

We interpret the mechanism underlying the probe beam amplification as Bragg scattering of pump photons into the probe beam direction via an atomic density grating.  The forces resulting from the interference of a pump and nearly counterpropagating probe beam generate an atomic density modulation along $\vec{G}=\vec{k}_{p}+\vec{k}_{s}$ that is phase-matched for Bragg scattering. The resulting nonlinear atomic polarization is $\vec{P}^{NL}=\eta \mu^2 b \vec{E}_p/\hbar \Delta$, where $b$ is the amplitude of the density modulation along $\hat{G}$ and $\vec{E}_p$ is the pump electric field \cite{piovella01}. Thermal motion limits the degree to which atoms can be localized along $\hat{G}$ so that lower temperatures result in larger values of $b$ and, consequently, higher Bragg scattering efficiencies.

Because of this connection between atomic temperature and NLO scattering efficiency, we find that the dissipative optical lattice formed by the pump beams plays an important  role in the nonlinearity (in contrast to other works in which the pump-beam lattice does not directly enhance the light-matter coupling \cite{chan03, baumann10}). The pump-beam lattice causes Sisyphus cooling, which transforms the atomic momentum distribution along $\hat{k}_p$ from a Maxwell-Boltzmann distribution to one that is well-described by a double-Gaussian function \cite{jersblad04, dion05}.  One interprets the gas as a non-thermal system consisting of a cold  component of atoms well-localized in the pump-pump lattice at a temperature $T'_c$ and an unbound, hot component at temperature $T'_h$ undergoing anomalous diffusion.  The cooling process transfers atoms from the hot to the cold component, where the fraction of atoms in each component is $f_{h,c}$, respectively.  In our experiment, we find that $T'_c\cong2.5$ $\mu$K and $T'_h\cong25$ $\mu$K and that both are largely insensitive to $I_p$ and $\Delta$ \cite{greenberg11}. 

The pump-beam lattice assists in cooling and loading atoms into the lattice along $\hat{G}$, which is phase-matched for scattering pump light into the probe beam directions. For weak bunching and $(2 \Delta/\Gamma)^2 \gg 1$, we find that \cite{saffman08}
\begin{equation}
\label{eq:b2}
b=\frac{\hbar \Gamma}{4 (\Delta/\Gamma)}\frac{\epsilon_0 c (E_s E^*_p+E^*_i E_p)}{I_{sat}}\left(\frac{f_c}{k_B T_c}+\frac{f_h}{k_B T_h} \right),
\end{equation}
where $E_s$ ($E_i$) is the  signal (idler) electric field strength and $T_{c,h}$ are the temperatures along $\hat{G}$ (which are approximately equal to $T'_{c,h}$ for $\theta/2 \ll 1$). We solve for $f_{c,h}$ by  calculating numerically the steady-state momentum distribution for a $J_g=1/2\rightarrow J_e=3/2$ transition  using a Bloch state approach \cite{castin91} and find that $f_c$ initially increases linearly with $I_p$ before asymptoting to 1.  While no analytic solution for $f_c$ exists,  the simple functional form $f_c=1-f_h \cong \text{tanh}(I_p/I_c)$ fits the numerical results well, where the characteristic intensity  for the cooling process (and, therefore, the nonlinearity) is $I_c$.  We find that $I_c$ is directly proportional to and of the same order as the  so-called d\'ecrochage intensity $I_d$, which is the intensity where the root mean squared width of the momentum distribution is a minimum (\textit{i.e.}, where cooling optimally balances diffusive heating) \cite{dalibard89}.  Perhaps surprisingly, both the cooling and heating rates vary with detuning in such a way that $I_d$ and, therefore, $I_c$ is independent of $\Delta$ \cite{jersblad00}.  As we discuss later, the independence of $I_c$ on $\Delta$ plays an important role in achieving large NLO coupling strengths in our system.

Substituting the atomic polarization given above into Maxwell's equations, we find that the steady-state coupled wave equations for the signal and idler beams become
\begin{equation}
\label{eq:cwe}
\frac{dE_s}{dz} = i \kappa E_s+i\beta E^*_i, 
\hspace{10pt} 
\frac{dE^*_i}{dz} =i \kappa^* E^*_i+i\beta E_s,
\end{equation}
where we make the constant-pump-beam approximation and assume that the optical fields adiabatically follow the evolution of the atomic density grating.  We define the nonlinear coupling coefficient  
\begin{equation}
\label{eq:beta}
\beta=\beta_c+\beta_h=\frac{\alpha_{\Delta=0} \hbar \Gamma}{64 (\Delta/\Gamma)^2}\frac{I_p}{I_{sat}} \left(\frac{f_c}{k_B T_c} + \frac{f_h}{k_B T_h} \right),
\end{equation}
 such that $\beta L$ corresponds to the nonlinear phase shift imposed on the weak probe beams by the interaction and $\kappa=\kappa_R+i \kappa_I =\beta+\alpha_{\Delta=0} (-2 \Delta/\Gamma+i)/8 (\Delta/\Gamma)^2$.

From Eq. \ref{eq:cwe}, we find that \cite{abrams78}
\begin{eqnarray}
\label{eq:soltns}
R=\frac{|\beta \text{sin}(w L)|^2}{|w \text{cos}(w L) + \kappa_I \text{sin}(w L)|},
\end{eqnarray}
where $w=(|\beta|^2-\kappa_I^2)^{1/2}$.  The solid lines in Fig. \ref{fig:scaling} demonstrate that the predictions of Eq. \ref{eq:soltns} fit the experimental data well ($\chi^2_{red}=1.6$) over all measured values of $\Delta$ and $I_p$.  Using the measured values of  $\alpha_{\Delta=0}L$, $\Delta$ and $T_{c,h}$, the only free parameter used to obtain this fit is $I_c$, which we assume is independent of $\Delta$.  We find that $I_c=1.5(\pm 0.5)$  mW/cm$^2$.  We can therefore access NLO interaction strengths comparable to those of resonantly-driven atoms while the atoms are nearly transparent, since $I_c\cong I_{sat}$ even for $|\Delta/\Gamma|\gg 1$.  

In order to quantify the nonlinear response, we use Eq. \ref{eq:soltns} to determine $\beta$ directly (see Fig. \ref{fig:scaling}c).  For small pump intensities (\textit{i.e.}, $I_p \ll I_c$), a Taylor series expansion of $\beta$ about $I_p=0$ yields  
\begin{equation}
\label{eq:betaT}
\beta\cong\frac{\alpha_{\Delta=0} \hbar \Gamma}{64 (\Delta/\Gamma)^2} \frac{I_p}{I_{sat}} \left(\frac{I_p/I_c}{k_B T_c} + \frac{1-I_p/I_c}{k_B T_h} \right).
\end{equation}
The quadratic (linear) components of $\beta$ correspond to a $\chi^{(5)}$ ($\chi^{(3)}$) response according to $\beta = |k_p|/(4 L \epsilon_0 c)[ \chi^{(3)}_{xxyy} I_p+(2 \epsilon_0 c)^{-1}\chi^{(5)}_{xxxxyy} I_p^2]$ for the polarizations shown in Fig. \ref{fig:setup}. Using this relationship, we find  $\chi^{(5)}_{xxxxyy}= 1.9(\pm 0.3)\times10^{-12}$ (m/V)$^4$  for $\Delta=-3\Gamma$ (see inset in Fig. \ref{fig:scaling}c).  This  $\chi^{(5)}$ response is due almost entirely to the cold component (since $T_c/T_h = 0.1$) and gives rise to six-wave mixing (SWM), where the pump beams simultaneously undergo Bragg scattering and transfer atoms from the hot to the cold component (\textit{i.e.}, cool the atoms).  The value of $\chi^{(5)}$ is the largest ever reported, exceeding that obtained for EIT-based SWM by $10^7$ \cite{zhang09, michinel06} and in three-photon absorption in a zinc blend semiconductor by $10^{24}$ \cite{cirloganu08}. 

For the same detuning ($\Delta=-3\Gamma$), we find $\chi^{(3)}_{xxyy}=2.0(\pm 2.3)\times 10^{-10}$ (m/V)$^2$, where the large experimental uncertainty arises from our inability to accurately measure $\beta$ at the smallest $I_p$.  This $\chi^{(3)}$ response is $\sim100$ times smaller than that observed in EIT-based systems \cite{lo11}, and  leads to four-wave mixing (FWM) in which the pump beams scatter off the weak atomic density grating composed solely of atoms in the hot component \cite{hemmerich94}. 
 
Beyond the region in which the Taylor series expansion given in Eq. \ref{eq:betaT} is valid (\textit{i.e.}, $I_c<I_p \ll I^\Delta_{sat}$), $f_c\cong1$ and the $\chi^{(5)}$ response saturates.  Nevertheless, $\beta$ continues to increase linearly with $I_p$, in contrast to $\chi^{(3)}$ processes that increase only sub-linearly beyond saturation.   

We compare our observed NLO response to previously-reported nonlinearities based on $\chi^{(3)}$ processes by considering the achievable NLO phase shift for a fixed $I_p$.  For example, Lo \textit{et al.} \cite{lo11} observe a 0.25 rad phase shift for an intensity of 230 $\mu$W/cm$^2$. We find that $\beta L=0.25$ rad for $I_p=560$ $\mu$W/cm$^2$ and $\Delta/\Gamma=-3$, although our system does not require any auxiliary, strong coupling beams (as in the EIT-based setups).  Also, $\beta$ continues increasing quadratically with $I_p$ for our $\chi^{(5)}$ process (rather than linearly as in $\chi^{(3)}$ processes), which further aides us in achieving large phase shifts at low-light-levels. Our system provides the additional advantage that we can work far from resonance and thereby combine large nonlinearities with high transparency (similar to recent predictions for Rydberg-enhanced EIT \cite{sevincli11}).  We quantify the tradeoff between absorptive loss and NLO phase shift using the  cross-phase-modulation figure of merit $\zeta \equiv \beta /\alpha_\Delta$  (\textit{i.e.}, the ratio of the NLO phase shift to power loss, see Fig. \ref{fig:scaling}d).  While Lo \textit{et al.}  \cite{lo11} observe a maximum value of $\zeta=0.35$ for an intensity of 230 $\mu$W/cm$^2$, we exceed this value for $I_p>300$ $\mu$W/cm$^2$ and obtain a maximum value of $\zeta=26$ for $I_p=5.6$ mW/cm$^2$. Our system's ability to realize large $\zeta$ arises directly from the independence of $I_c$ on detuning.

\begin{figure}[b, h, t]
 \includegraphics[width=3.4in]{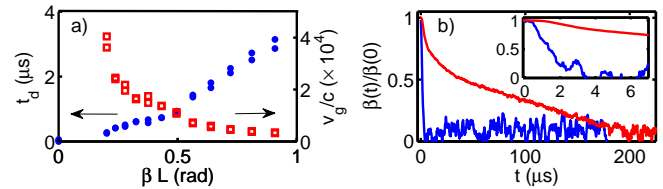}
 \caption{ a) Dependence of the group velocity and slow light delay  on the NLO coupling strength. b) Decay of the nonlinear coupling coefficient as a function of time after the signal beam is turned off at $t=0$ for $\Delta=-5 \Gamma$.  We measure the long (short) decay time of 100 (1.8) $\mu$s using $I_p$= 1.25 (0.45) mW/cm$^2$, which corresponds to $\beta L$= 0.84 (0.11).  The inset shows a detailed view at small times.}
 \label{fig:decay}
 \end{figure}

For sufficiently large values of $\beta L$, the back-action of the amplified probe fields  influences the coupled light-matter dynamics, resulting in a strongly-coupled system with long-range atom-atom interactions.  In this regime, additional cooling occurs in the $x$-$z$-plane  due to the lattice formed by the amplified probe and nearly-\textit{copropagating} pump beams \cite{chan03,greenberg11}.  We also observe small group velocities of light $v_g \cong c/10^5$ (corresponding to a slow light delay of $t_d\cong 4 $ $\mu$s, see Fig. \ref{fig:decay}a) by weakly modulating $I_s$ and recording the phase lag of the transmitted beam (relative to the case where $I_p=0$).  The medium therefore acts like a high-finesse cavity and lengthens the photon lifetime in the gas such that the time scales of the atomic and optical dynamics become comparable (contrary to the assumptions leading to Eq. \ref{eq:cwe}).  As a result of this back-action at large $\beta L$, the atomic coherence time (defined as the $1/e$ decay time) increases to over $100$ $\mu$s, which is over 50 times larger than that measured for $\beta L \ll 1$, where thermal atomic motion causes grating washout (see Fig. \ref{fig:decay}b).  In addition, we observe a collective instability for $\beta L>1$ in which the probe fields and density grating are spontaneously self-generated via the NLO interaction.  We will describe
these results in a future publication.   

Our results are not limited to the particular experimental geometry.  We observe qualitatively similar results for other pump field polarizations (\textit{e.g.}, lin$\theta$lin and $\sigma^+$-$\sigma^+$), which indicates that the details of the configuration are not essential to the nonlinearity.  On the other hand, we find that $\beta\sim100\times$ smaller for $\Delta>0$, where heating and limited bunching occur.  Thus, the NLO process occurs for any setup that permits both atomic cooling and bunching, which is consistent with our physical interpretation of the nonlinearity.

We gratefully acknowledge the financial support of the NSF through Grant \#PHY-0855399.


\bibliography{chi5_opex_bibtex}

\begin{thebibliography}{35}%
\makeatletter
\providecommand \@ifxundefined [1]{%
 \@ifx{#1\undefined}
}%
\providecommand \@ifnum [1]{%
 \ifnum #1\expandafter \@firstoftwo
 \else \expandafter \@secondoftwo
 \fi
}%
\providecommand \@ifx [1]{%
 \ifx #1\expandafter \@firstoftwo
 \else \expandafter \@secondoftwo
 \fi
}%
\providecommand \natexlab [1]{#1}%
\providecommand \enquote  [1]{``#1''}%
\providecommand \bibnamefont  [1]{#1}%
\providecommand \bibfnamefont [1]{#1}%
\providecommand \citenamefont [1]{#1}%
\providecommand \href@noop [0]{\@secondoftwo}%
\providecommand \href [0]{\begingroup \@sanitize@url \@href}%
\providecommand \@href[1]{\@@startlink{#1}\@@href}%
\providecommand \@@href[1]{\endgroup#1\@@endlink}%
\providecommand \@sanitize@url [0]{\catcode `\\12\catcode `\$12\catcode
  `\&12\catcode `\#12\catcode `\^12\catcode `\_12\catcode `\%12\relax}%
\providecommand \@@startlink[1]{}%
\providecommand \@@endlink[0]{}%
\providecommand \url  [0]{\begingroup\@sanitize@url \@url }%
\providecommand \@url [1]{\endgroup\@href {#1}{\urlprefix }}%
\providecommand \urlprefix  [0]{URL }%
\providecommand \Eprint [0]{\href }%
\providecommand \doibase [0]{http://dx.doi.org/}%
\providecommand \selectlanguage [0]{\@gobble}%
\providecommand \bibinfo  [0]{\@secondoftwo}%
\providecommand \bibfield  [0]{\@secondoftwo}%
\providecommand \translation [1]{[#1]}%
\providecommand \BibitemOpen [0]{}%
\providecommand \bibitemStop [0]{}%
\providecommand \bibitemNoStop [0]{.\EOS\space}%
\providecommand \EOS [0]{\spacefactor3000\relax}%
\providecommand \BibitemShut  [1]{\csname bibitem#1\endcsname}%
\let\auto@bib@innerbib\@empty
\bibitem [{\citenamefont {Kimble}(2008)}]{kimble08}%
  \BibitemOpen
  \bibfield  {author} {\bibinfo {author} {\bibfnamefont {H.~J.}\ \bibnamefont
  {Kimble}},\ }\href@noop {} {\bibfield  {journal} {\bibinfo  {journal}
  {Nature}\ }\textbf {\bibinfo {volume} {453}},\ \bibinfo {pages} {1023}
  (\bibinfo {year} {2008})}\BibitemShut {NoStop}%
\bibitem [{\citenamefont {Boyd}(2008)}]{boydnlo}%
  \BibitemOpen
  \bibfield  {author} {\bibinfo {author} {\bibfnamefont {R.~W.}\ \bibnamefont
  {Boyd}},\ }\enquote {\bibinfo {title} {Nonlinear optics},}\ \ (\bibinfo
  {publisher} {Academic Press},\ \bibinfo {year} {2008})\ Chap.~\bibinfo
  {chapter} {3},\ \bibinfo {edition} {3rd}\ ed.\BibitemShut {Stop}%
\bibitem [{\citenamefont {Hau}\ \emph {et~al.}(1999)\citenamefont {Hau},
  \citenamefont {Harris}, \citenamefont {Dutton},\ and\ \citenamefont
  {Behroozi}}]{hau99}%
  \BibitemOpen
  \bibfield  {author} {\bibinfo {author} {\bibfnamefont {L.~V.}\ \bibnamefont
  {Hau}}, \bibinfo {author} {\bibfnamefont {S.~E.}\ \bibnamefont {Harris}},
  \bibinfo {author} {\bibfnamefont {Z.}~\bibnamefont {Dutton}}, \ and\ \bibinfo
  {author} {\bibfnamefont {C.~H.}\ \bibnamefont {Behroozi}},\ }\href@noop {}
  {\bibfield  {journal} {\bibinfo  {journal} {Nature}\ }\textbf {\bibinfo
  {volume} {397}},\ \bibinfo {pages} {594} (\bibinfo {year}
  {1999})}\BibitemShut {NoStop}%
\bibitem [{\citenamefont {Shiau}\ \emph {et~al.}(2011)\citenamefont {Shiau},
  \citenamefont {Wu}, \citenamefont {Lin},\ and\ \citenamefont
  {Chen}}]{shiau11}%
  \BibitemOpen
  \bibfield  {author} {\bibinfo {author} {\bibfnamefont {B.-W.}\ \bibnamefont
  {Shiau}}, \bibinfo {author} {\bibfnamefont {M.-C.}\ \bibnamefont {Wu}},
  \bibinfo {author} {\bibfnamefont {C.-C.}\ \bibnamefont {Lin}}, \ and\
  \bibinfo {author} {\bibfnamefont {Y.-C.}\ \bibnamefont {Chen}},\ }\href@noop
  {} {\bibfield  {journal} {\bibinfo  {journal} {Phys. Rev. Lett.}\ }\textbf
  {\bibinfo {volume} {106}},\ \bibinfo {pages} {193006} (\bibinfo {year}
  {2011})}\BibitemShut {NoStop}%
\bibitem [{\citenamefont {Sevin\ifmmode~\mbox{\c{c}}\else \c{c}\fi{}li}\ \emph
  {et~al.}(2011)\citenamefont {Sevin\ifmmode~\mbox{\c{c}}\else \c{c}\fi{}li},
  \citenamefont {Henkel}, \citenamefont {Ates},\ and\ \citenamefont
  {Pohl}}]{sevincli11}%
  \BibitemOpen
  \bibfield  {author} {\bibinfo {author} {\bibfnamefont {S.}~\bibnamefont
  {Sevin\ifmmode~\mbox{\c{c}}\else \c{c}\fi{}li}}, \bibinfo {author}
  {\bibfnamefont {N.}~\bibnamefont {Henkel}}, \bibinfo {author} {\bibfnamefont
  {C.}~\bibnamefont {Ates}}, \ and\ \bibinfo {author} {\bibfnamefont
  {T.}~\bibnamefont {Pohl}},\ }\href@noop {} {\bibfield  {journal} {\bibinfo
  {journal} {Phys. Rev. Lett.}\ }\textbf {\bibinfo {volume} {107}},\ \bibinfo
  {pages} {153001} (\bibinfo {year} {2011})}\BibitemShut {NoStop}%
\bibitem [{\citenamefont {Lo}\ \emph {et~al.}(2011)\citenamefont {Lo},
  \citenamefont {Chen}, \citenamefont {Su}, \citenamefont {Chen}, \citenamefont
  {Chen}, \citenamefont {Chen}, \citenamefont {Yu},\ and\ \citenamefont
  {Chen}}]{lo11}%
  \BibitemOpen
  \bibfield  {author} {\bibinfo {author} {\bibfnamefont {H.-Y.}\ \bibnamefont
  {Lo}}, \bibinfo {author} {\bibfnamefont {Y.-C.}\ \bibnamefont {Chen}},
  \bibinfo {author} {\bibfnamefont {P.-C.}\ \bibnamefont {Su}}, \bibinfo
  {author} {\bibfnamefont {H.-C.}\ \bibnamefont {Chen}}, \bibinfo {author}
  {\bibfnamefont {J.-X.}\ \bibnamefont {Chen}}, \bibinfo {author}
  {\bibfnamefont {Y.-C.}\ \bibnamefont {Chen}}, \bibinfo {author}
  {\bibfnamefont {I.~A.}\ \bibnamefont {Yu}}, \ and\ \bibinfo {author}
  {\bibfnamefont {Y.-F.}\ \bibnamefont {Chen}},\ }\href@noop {} {\bibfield
  {journal} {\bibinfo  {journal} {Phys. Rev. A}\ }\textbf {\bibinfo {volume}
  {83}},\ \bibinfo {pages} {041804(R)} (\bibinfo {year} {2011})}\BibitemShut
  {NoStop}%
\bibitem [{\citenamefont {Fleischhauer}\ \emph {et~al.}(2005)\citenamefont
  {Fleischhauer}, \citenamefont {Imamoglu},\ and\ \citenamefont
  {Marangos}}]{fleischhauer05}%
  \BibitemOpen
  \bibfield  {author} {\bibinfo {author} {\bibfnamefont {M.}~\bibnamefont
  {Fleischhauer}}, \bibinfo {author} {\bibfnamefont {A.}~\bibnamefont
  {Imamoglu}}, \ and\ \bibinfo {author} {\bibfnamefont {J.~P.}\ \bibnamefont
  {Marangos}},\ }\href@noop {} {\bibfield  {journal} {\bibinfo  {journal} {Rev.
  Mod. Phys.}\ }\textbf {\bibinfo {volume} {77}},\ \bibinfo {pages} {633}
  (\bibinfo {year} {2005})}\BibitemShut {NoStop}%
\bibitem [{\citenamefont {Venkataraman}\ \emph {et~al.}(2011)\citenamefont
  {Venkataraman}, \citenamefont {Saha}, \citenamefont {Londero},\ and\
  \citenamefont {Gaeta}}]{venkataraman11}%
  \BibitemOpen
  \bibfield  {author} {\bibinfo {author} {\bibfnamefont {V.}~\bibnamefont
  {Venkataraman}}, \bibinfo {author} {\bibfnamefont {K.}~\bibnamefont {Saha}},
  \bibinfo {author} {\bibfnamefont {P.}~\bibnamefont {Londero}}, \ and\
  \bibinfo {author} {\bibfnamefont {A.~L.}\ \bibnamefont {Gaeta}},\ }\href@noop
  {} {\bibfield  {journal} {\bibinfo  {journal} {Phys. Rev. Lett.}\ }\textbf
  {\bibinfo {volume} {107}},\ \bibinfo {pages} {193902} (\bibinfo {year}
  {2011})}\BibitemShut {NoStop}%
\bibitem [{\citenamefont {Dawes}\ \emph {et~al.}(2010)\citenamefont {Dawes},
  \citenamefont {Gauthier}, \citenamefont {Schumacher}, \citenamefont {Kwong},
  \citenamefont {Binder},\ and\ \citenamefont {Smirl}}]{dawes10}%
  \BibitemOpen
  \bibfield  {author} {\bibinfo {author} {\bibfnamefont {A.~M.~C.}\
  \bibnamefont {Dawes}}, \bibinfo {author} {\bibfnamefont {D.~J.}\ \bibnamefont
  {Gauthier}}, \bibinfo {author} {\bibfnamefont {S.}~\bibnamefont
  {Schumacher}}, \bibinfo {author} {\bibfnamefont {N.~H.}\ \bibnamefont
  {Kwong}}, \bibinfo {author} {\bibfnamefont {R.}~\bibnamefont {Binder}}, \
  and\ \bibinfo {author} {\bibfnamefont {A.}~\bibnamefont {Smirl}},\
  }\href@noop {} {\bibfield  {journal} {\bibinfo  {journal} {Laser and Photon.
  Rev.}\ }\textbf {\bibinfo {volume} {4}},\ \bibinfo {pages} {221} (\bibinfo
  {year} {2010})}\BibitemShut {NoStop}%
\bibitem [{\citenamefont {Michinel}\ \emph {et~al.}(2006)\citenamefont
  {Michinel}, \citenamefont {Paz-Alonso},\ and\ \citenamefont
  {P{\'e}rez-Garc{\'i}a}}]{michinel06}%
  \BibitemOpen
  \bibfield  {author} {\bibinfo {author} {\bibfnamefont {H.}~\bibnamefont
  {Michinel}}, \bibinfo {author} {\bibfnamefont {M.~J.}\ \bibnamefont
  {Paz-Alonso}}, \ and\ \bibinfo {author} {\bibfnamefont {V.~M.}\ \bibnamefont
  {P{\'e}rez-Garc{\'i}a}},\ }\href@noop {} {\bibfield  {journal} {\bibinfo
  {journal} {Phys. Rev. Lett.}\ }\textbf {\bibinfo {volume} {96}},\ \bibinfo
  {pages} {023903} (\bibinfo {year} {2006})}\BibitemShut {NoStop}%
\bibitem [{\citenamefont {Fibich}\ \emph {et~al.}(2007)\citenamefont {Fibich},
  \citenamefont {Gavish},\ and\ \citenamefont {Wang}}]{fibich07}%
  \BibitemOpen
  \bibfield  {author} {\bibinfo {author} {\bibfnamefont {G.}~\bibnamefont
  {Fibich}}, \bibinfo {author} {\bibfnamefont {N.}~\bibnamefont {Gavish}}, \
  and\ \bibinfo {author} {\bibfnamefont {X.~P.}\ \bibnamefont {Wang}},\
  }\href@noop {} {\bibfield  {journal} {\bibinfo  {journal} {Physica D}\
  }\textbf {\bibinfo {volume} {231}},\ \bibinfo {pages} {55} (\bibinfo {year}
  {2007})}\BibitemShut {NoStop}%
\bibitem [{\citenamefont {Hang}\ \emph {et~al.}(2006)\citenamefont {Hang},
  \citenamefont {Li}, \citenamefont {Ma},\ and\ \citenamefont
  {Huang}}]{hang06}%
  \BibitemOpen
  \bibfield  {author} {\bibinfo {author} {\bibfnamefont {C.}~\bibnamefont
  {Hang}}, \bibinfo {author} {\bibfnamefont {Y.}~\bibnamefont {Li}}, \bibinfo
  {author} {\bibfnamefont {L.}~\bibnamefont {Ma}}, \ and\ \bibinfo {author}
  {\bibfnamefont {G.}~\bibnamefont {Huang}},\ }\href@noop {} {\bibfield
  {journal} {\bibinfo  {journal} {Phys. Rev. A}\ }\textbf {\bibinfo {volume}
  {74}},\ \bibinfo {pages} {012319} (\bibinfo {year} {2006})}\BibitemShut
  {NoStop}%
\bibitem [{\citenamefont {Zubairy}\ \emph {et~al.}(2002)\citenamefont
  {Zubairy}, \citenamefont {Matsko},\ and\ \citenamefont {Scully}}]{zubairy02}%
  \BibitemOpen
  \bibfield  {author} {\bibinfo {author} {\bibfnamefont {M.~S.}\ \bibnamefont
  {Zubairy}}, \bibinfo {author} {\bibfnamefont {A.~B.}\ \bibnamefont {Matsko}},
  \ and\ \bibinfo {author} {\bibfnamefont {M.~O.}\ \bibnamefont {Scully}},\
  }\href@noop {} {\bibfield  {journal} {\bibinfo  {journal} {Phys. Rev. A.}\
  }\textbf {\bibinfo {volume} {65}},\ \bibinfo {pages} {043804} (\bibinfo
  {year} {2002})}\BibitemShut {NoStop}%
\bibitem [{\citenamefont {Felinto}\ \emph {et~al.}(2010)\citenamefont
  {Felinto}, \citenamefont {Moretti}, \citenamefont {de~Oliveira},\ and\
  \citenamefont {Tabosa}}]{felinto10}%
  \BibitemOpen
  \bibfield  {author} {\bibinfo {author} {\bibfnamefont {D.}~\bibnamefont
  {Felinto}}, \bibinfo {author} {\bibfnamefont {D.}~\bibnamefont {Moretti}},
  \bibinfo {author} {\bibfnamefont {R.}~\bibnamefont {de~Oliveira}}, \ and\
  \bibinfo {author} {\bibfnamefont {J.}~\bibnamefont {Tabosa}},\ }\href@noop {}
  {\bibfield  {journal} {\bibinfo  {journal} {Opt. Lett.}\ }\textbf {\bibinfo
  {volume} {35}},\ \bibinfo {pages} {3937} (\bibinfo {year}
  {2010})}\BibitemShut {NoStop}%
\bibitem [{\citenamefont {Kang}\ \emph {et~al.}(2004)\citenamefont {Kang},
  \citenamefont {Hernandez},\ and\ \citenamefont {Zhu}}]{kang04}%
  \BibitemOpen
  \bibfield  {author} {\bibinfo {author} {\bibfnamefont {H.}~\bibnamefont
  {Kang}}, \bibinfo {author} {\bibfnamefont {G.}~\bibnamefont {Hernandez}}, \
  and\ \bibinfo {author} {\bibfnamefont {Y.}~\bibnamefont {Zhu}},\ }\href@noop
  {} {\bibfield  {journal} {\bibinfo  {journal} {Phys. Rev. Lett.}\ }\textbf
  {\bibinfo {volume} {93}},\ \bibinfo {pages} {073601} (\bibinfo {year}
  {2004})}\BibitemShut {NoStop}%
\bibitem [{\citenamefont {Zhang}\ \emph {et~al.}(2009)\citenamefont {Zhang},
  \citenamefont {Khadka}, \citenamefont {Anderson},\ and\ \citenamefont
  {Xiao}}]{zhang09}%
  \BibitemOpen
  \bibfield  {author} {\bibinfo {author} {\bibfnamefont {Y.}~\bibnamefont
  {Zhang}}, \bibinfo {author} {\bibfnamefont {U.}~\bibnamefont {Khadka}},
  \bibinfo {author} {\bibfnamefont {B.}~\bibnamefont {Anderson}}, \ and\
  \bibinfo {author} {\bibfnamefont {M.}~\bibnamefont {Xiao}},\ }\href@noop {}
  {\bibfield  {journal} {\bibinfo  {journal} {Phys. Rev. Lett.}\ }\textbf
  {\bibinfo {volume} {102}},\ \bibinfo {pages} {013601} (\bibinfo {year}
  {2009})}\BibitemShut {NoStop}%
\bibitem [{\citenamefont {Giri}\ and\ \citenamefont {Gupta}(2003)}]{giri03}%
  \BibitemOpen
  \bibfield  {author} {\bibinfo {author} {\bibfnamefont {D.~K.}\ \bibnamefont
  {Giri}}\ and\ \bibinfo {author} {\bibfnamefont {P.}~\bibnamefont {Gupta}},\
  }\href@noop {} {\bibfield  {journal} {\bibinfo  {journal} {Optics
  Communications}\ }\textbf {\bibinfo {volume} {221}},\ \bibinfo {pages} {135 }
  (\bibinfo {year} {2003})}\BibitemShut {NoStop}%
\bibitem [{\citenamefont {Dalibard}\ and\ \citenamefont
  {Cohen-Tannoudji}(1989)}]{dalibard89}%
  \BibitemOpen
  \bibfield  {author} {\bibinfo {author} {\bibfnamefont {J.}~\bibnamefont
  {Dalibard}}\ and\ \bibinfo {author} {\bibfnamefont {C.}~\bibnamefont
  {Cohen-Tannoudji}},\ }\href@noop {} {\bibfield  {journal} {\bibinfo
  {journal} {J. Opt. Soc. Am. B}\ }\textbf {\bibinfo {volume} {6}},\ \bibinfo
  {pages} {2023} (\bibinfo {year} {1989})}\BibitemShut {NoStop}%
\bibitem [{\citenamefont {Inouye}\ \emph {et~al.}(1999)\citenamefont {Inouye},
  \citenamefont {Chikkatur}, \citenamefont {Stamper-Kurn}, \citenamefont
  {Stenger}, \citenamefont {Pritchard},\ and\ \citenamefont
  {Ketterle}}]{inouye99}%
  \BibitemOpen
  \bibfield  {author} {\bibinfo {author} {\bibfnamefont {S.}~\bibnamefont
  {Inouye}}, \bibinfo {author} {\bibfnamefont {A.~P.}\ \bibnamefont
  {Chikkatur}}, \bibinfo {author} {\bibfnamefont {D.~M.}\ \bibnamefont
  {Stamper-Kurn}}, \bibinfo {author} {\bibfnamefont {J.}~\bibnamefont
  {Stenger}}, \bibinfo {author} {\bibfnamefont {D.~E.}\ \bibnamefont
  {Pritchard}}, \ and\ \bibinfo {author} {\bibfnamefont {W.}~\bibnamefont
  {Ketterle}},\ }\href@noop {} {\bibfield  {journal} {\bibinfo  {journal}
  {Science}\ }\textbf {\bibinfo {volume} {285}},\ \bibinfo {pages} {571}
  (\bibinfo {year} {1999})}\BibitemShut {NoStop}%
\bibitem [{\citenamefont {Kruse}\ \emph {et~al.}(2003)\citenamefont {Kruse},
  \citenamefont {von Cube}, \citenamefont {Zimmermann},\ and\ \citenamefont
  {Courteille}}]{kruse03}%
  \BibitemOpen
  \bibfield  {author} {\bibinfo {author} {\bibfnamefont {D.}~\bibnamefont
  {Kruse}}, \bibinfo {author} {\bibfnamefont {C.}~\bibnamefont {von Cube}},
  \bibinfo {author} {\bibfnamefont {C.}~\bibnamefont {Zimmermann}}, \ and\
  \bibinfo {author} {\bibfnamefont {P.~W.}\ \bibnamefont {Courteille}},\
  }\href@noop {} {\bibfield  {journal} {\bibinfo  {journal} {Phys. Rev. Lett.}\
  }\textbf {\bibinfo {volume} {91}},\ \bibinfo {pages} {183601} (\bibinfo
  {year} {2003})}\BibitemShut {NoStop}%
\bibitem [{\citenamefont {Schilke}\ \emph {et~al.}(2011)\citenamefont
  {Schilke}, \citenamefont {Zimmermann}, \citenamefont {Courteille},\ and\
  \citenamefont {Guerin}}]{schilke11}%
  \BibitemOpen
  \bibfield  {author} {\bibinfo {author} {\bibfnamefont {A.}~\bibnamefont
  {Schilke}}, \bibinfo {author} {\bibfnamefont {C.}~\bibnamefont {Zimmermann}},
  \bibinfo {author} {\bibfnamefont {P.}~\bibnamefont {Courteille}}, \ and\
  \bibinfo {author} {\bibfnamefont {W.}~\bibnamefont {Guerin}},\ }\href@noop {}
  {\bibfield  {journal} {\bibinfo  {journal} {Opt. Lett.}\ }\textbf {\bibinfo
  {volume} {106}},\ \bibinfo {pages} {223903} (\bibinfo {year}
  {2011})}\BibitemShut {NoStop}%
\bibitem [{\citenamefont {Gopalakrishnan}\ \emph {et~al.}(2010)\citenamefont
  {Gopalakrishnan}, \citenamefont {Lev},\ and\ \citenamefont
  {Goldbart}}]{gopal10}%
  \BibitemOpen
  \bibfield  {author} {\bibinfo {author} {\bibfnamefont {S.}~\bibnamefont
  {Gopalakrishnan}}, \bibinfo {author} {\bibfnamefont {B.~L.}\ \bibnamefont
  {Lev}}, \ and\ \bibinfo {author} {\bibfnamefont {P.~M.}\ \bibnamefont
  {Goldbart}},\ }\href@noop {} {\bibfield  {journal} {\bibinfo  {journal}
  {Phys. Rev. A}\ }\textbf {\bibinfo {volume} {82}},\ \bibinfo {pages} {043612}
  (\bibinfo {year} {2010})}\BibitemShut {NoStop}%
\bibitem [{\citenamefont {Greenberg}\ \emph {et~al.}(2007)\citenamefont
  {Greenberg}, \citenamefont {Ori\'a}, \citenamefont {Dawes},\ and\
  \citenamefont {Gauthier}}]{greenberg07}%
  \BibitemOpen
  \bibfield  {author} {\bibinfo {author} {\bibfnamefont {J.~A.}\ \bibnamefont
  {Greenberg}}, \bibinfo {author} {\bibfnamefont {M.}~\bibnamefont {Ori\'a}},
  \bibinfo {author} {\bibfnamefont {A.~M.~C.}\ \bibnamefont {Dawes}}, \ and\
  \bibinfo {author} {\bibfnamefont {D.~J.}\ \bibnamefont {Gauthier}},\
  }\href@noop {} {\bibfield  {journal} {\bibinfo  {journal} {Opt. Exp.}\
  }\textbf {\bibinfo {volume} {15}},\ \bibinfo {pages} {17699} (\bibinfo {year}
  {2007})}\BibitemShut {NoStop}%
\bibitem [{\citenamefont {Greenberg}\ \emph {et~al.}(2011)\citenamefont
  {Greenberg}, \citenamefont {Schmittberger},\ and\ \citenamefont
  {Gauthier}}]{greenberg11}%
  \BibitemOpen
  \bibfield  {author} {\bibinfo {author} {\bibfnamefont {J.~A.}\ \bibnamefont
  {Greenberg}}, \bibinfo {author} {\bibfnamefont {B.~L.}\ \bibnamefont
  {Schmittberger}}, \ and\ \bibinfo {author} {\bibfnamefont {D.~J.}\
  \bibnamefont {Gauthier}},\ }\href@noop {} {\bibfield  {journal} {\bibinfo
  {journal} {Opt. Express}\ }\textbf {\bibinfo {volume} {19}},\ \bibinfo
  {pages} {22535} (\bibinfo {year} {2011})}\BibitemShut {NoStop}%
\bibitem [{\citenamefont {Piovella}\ \emph {et~al.}(2001)\citenamefont
  {Piovella}, \citenamefont {Bonifacio}, \citenamefont {McNeil},\ and\
  \citenamefont {Robb}}]{piovella01}%
  \BibitemOpen
  \bibfield  {author} {\bibinfo {author} {\bibfnamefont {N.}~\bibnamefont
  {Piovella}}, \bibinfo {author} {\bibfnamefont {R.}~\bibnamefont {Bonifacio}},
  \bibinfo {author} {\bibfnamefont {B.~W.~J.}\ \bibnamefont {McNeil}}, \ and\
  \bibinfo {author} {\bibfnamefont {G.~R.~M.}\ \bibnamefont {Robb}},\
  }\href@noop {} {\bibfield  {journal} {\bibinfo  {journal} {Opt. Comm.}\
  }\textbf {\bibinfo {volume} {187}},\ \bibinfo {pages} {165} (\bibinfo {year}
  {2001})}\BibitemShut {NoStop}%
\bibitem [{\citenamefont {Chan}\ \emph {et~al.}(2003)\citenamefont {Chan},
  \citenamefont {Black},\ and\ \citenamefont {Vuleti{\'c}}}]{chan03}%
  \BibitemOpen
  \bibfield  {author} {\bibinfo {author} {\bibfnamefont {H.~W.}\ \bibnamefont
  {Chan}}, \bibinfo {author} {\bibfnamefont {A.~T.}\ \bibnamefont {Black}}, \
  and\ \bibinfo {author} {\bibfnamefont {V.}~\bibnamefont {Vuleti{\'c}}},\
  }\href@noop {} {\bibfield  {journal} {\bibinfo  {journal} {Phys. Rev. Lett.}\
  }\textbf {\bibinfo {volume} {90}},\ \bibinfo {pages} {063003} (\bibinfo
  {year} {2003})}\BibitemShut {NoStop}%
\bibitem [{\citenamefont {Baumann}\ \emph {et~al.}(2010)\citenamefont
  {Baumann}, \citenamefont {Guerlin}, \citenamefont {Brennecke},\ and\
  \citenamefont {Esslinger}}]{baumann10}%
  \BibitemOpen
  \bibfield  {author} {\bibinfo {author} {\bibfnamefont {K.}~\bibnamefont
  {Baumann}}, \bibinfo {author} {\bibfnamefont {C.}~\bibnamefont {Guerlin}},
  \bibinfo {author} {\bibfnamefont {F.}~\bibnamefont {Brennecke}}, \ and\
  \bibinfo {author} {\bibfnamefont {T.}~\bibnamefont {Esslinger}},\ }\href@noop
  {} {\bibfield  {journal} {\bibinfo  {journal} {Nature}\ }\textbf {\bibinfo
  {volume} {464}},\ \bibinfo {pages} {1301 } (\bibinfo {year}
  {2010})}\BibitemShut {NoStop}%
\bibitem [{\citenamefont {Jersblad}\ \emph {et~al.}(2004)\citenamefont
  {Jersblad}, \citenamefont {Ellmann}, \citenamefont {Stochkel}, \citenamefont
  {Kastberg}, \citenamefont {Sanchez-Palencia},\ and\ \citenamefont
  {Kaiser}}]{jersblad04}%
  \BibitemOpen
  \bibfield  {author} {\bibinfo {author} {\bibfnamefont {J.}~\bibnamefont
  {Jersblad}}, \bibinfo {author} {\bibfnamefont {H.}~\bibnamefont {Ellmann}},
  \bibinfo {author} {\bibfnamefont {K.}~\bibnamefont {Stochkel}}, \bibinfo
  {author} {\bibfnamefont {A.}~\bibnamefont {Kastberg}}, \bibinfo {author}
  {\bibfnamefont {L.}~\bibnamefont {Sanchez-Palencia}}, \ and\ \bibinfo
  {author} {\bibfnamefont {R.}~\bibnamefont {Kaiser}},\ }\href@noop {}
  {\bibfield  {journal} {\bibinfo  {journal} {Phys. Rev. A}\ }\textbf {\bibinfo
  {volume} {69}},\ \bibinfo {pages} {013410} (\bibinfo {year}
  {2004})}\BibitemShut {NoStop}%
\bibitem [{\citenamefont {Dion}\ \emph {et~al.}(2005)\citenamefont {Dion},
  \citenamefont {Sjolund}, \citenamefont {Petra}, \citenamefont {Jonsell},\
  and\ \citenamefont {Kastberg}}]{dion05}%
  \BibitemOpen
  \bibfield  {author} {\bibinfo {author} {\bibfnamefont {C.}~\bibnamefont
  {Dion}}, \bibinfo {author} {\bibfnamefont {P.}~\bibnamefont {Sjolund}},
  \bibinfo {author} {\bibfnamefont {S.}~\bibnamefont {Petra}}, \bibinfo
  {author} {\bibfnamefont {S.}~\bibnamefont {Jonsell}}, \ and\ \bibinfo
  {author} {\bibfnamefont {A.}~\bibnamefont {Kastberg}},\ }\href@noop {}
  {\bibfield  {journal} {\bibinfo  {journal} {Eurphys. Lett.}\ }\textbf
  {\bibinfo {volume} {72}},\ \bibinfo {pages} {369} (\bibinfo {year}
  {2005})}\BibitemShut {NoStop}%
\bibitem [{\citenamefont {Saffman}\ and\ \citenamefont
  {Wang}(2008)}]{saffman08}%
  \BibitemOpen
  \bibfield  {author} {\bibinfo {author} {\bibfnamefont {M.}~\bibnamefont
  {Saffman}}\ and\ \bibinfo {author} {\bibfnamefont {Y.}~\bibnamefont {Wang}},\
  }in\ \href@noop {} {\emph {\bibinfo {booktitle} {Dissipative Solitons: From
  Optics to Biology and Medicine}}},\ \bibinfo {series} {Lecture Notes in
  Physics}, Vol.\ \bibinfo {volume} {751}\ (\bibinfo  {publisher} {Springer
  Berlin / Heidelberg},\ \bibinfo {year} {2008})\BibitemShut {NoStop}%
\bibitem [{\citenamefont {Castin}\ and\ \citenamefont
  {Dalibard}(1991)}]{castin91}%
  \BibitemOpen
  \bibfield  {author} {\bibinfo {author} {\bibfnamefont {Y.}~\bibnamefont
  {Castin}}\ and\ \bibinfo {author} {\bibfnamefont {J.}~\bibnamefont
  {Dalibard}},\ }\href@noop {} {\bibfield  {journal} {\bibinfo  {journal}
  {Europhys. Lett.}\ }\textbf {\bibinfo {volume} {14}},\ \bibinfo {pages} {761}
  (\bibinfo {year} {1991})}\BibitemShut {NoStop}%
\bibitem [{\citenamefont {Jersblad}\ \emph {et~al.}(2000)\citenamefont
  {Jersblad}, \citenamefont {Ellmann},\ and\ \citenamefont
  {Kastberg}}]{jersblad00}%
  \BibitemOpen
  \bibfield  {author} {\bibinfo {author} {\bibfnamefont {J.}~\bibnamefont
  {Jersblad}}, \bibinfo {author} {\bibfnamefont {H.}~\bibnamefont {Ellmann}}, \
  and\ \bibinfo {author} {\bibfnamefont {A.}~\bibnamefont {Kastberg}},\
  }\href@noop {} {\bibfield  {journal} {\bibinfo  {journal} {Phys. Rev. A}\
  }\textbf {\bibinfo {volume} {62}},\ \bibinfo {pages} {051401} (\bibinfo
  {year} {2000})}\BibitemShut {NoStop}%
\bibitem [{\citenamefont {Abrams}\ and\ \citenamefont {Lind}(1978)}]{abrams78}%
  \BibitemOpen
  \bibfield  {author} {\bibinfo {author} {\bibfnamefont {R.~L.}\ \bibnamefont
  {Abrams}}\ and\ \bibinfo {author} {\bibfnamefont {R.~C.}\ \bibnamefont
  {Lind}},\ }\href@noop {} {\bibfield  {journal} {\bibinfo  {journal} {Opt.
  Lett.}\ }\textbf {\bibinfo {volume} {2}},\ \bibinfo {pages} {94} (\bibinfo
  {year} {1978})}\BibitemShut {NoStop}%
\bibitem [{\citenamefont {Cirloganu}\ \emph {et~al.}(2008)\citenamefont
  {Cirloganu}, \citenamefont {Olszak}, \citenamefont {Padilha}, \citenamefont
  {Webster}, \citenamefont {Hagan},\ and\ \citenamefont
  {Stryland}}]{cirloganu08}%
  \BibitemOpen
  \bibfield  {author} {\bibinfo {author} {\bibfnamefont {C.~M.}\ \bibnamefont
  {Cirloganu}}, \bibinfo {author} {\bibfnamefont {P.~D.}\ \bibnamefont
  {Olszak}}, \bibinfo {author} {\bibfnamefont {L.~A.}\ \bibnamefont {Padilha}},
  \bibinfo {author} {\bibfnamefont {S.}~\bibnamefont {Webster}}, \bibinfo
  {author} {\bibfnamefont {D.~J.}\ \bibnamefont {Hagan}}, \ and\ \bibinfo
  {author} {\bibfnamefont {E.~W.~V.}\ \bibnamefont {Stryland}},\ }\href@noop {}
  {\bibfield  {journal} {\bibinfo  {journal} {Opt. Lett.}\ }\textbf {\bibinfo
  {volume} {33}},\ \bibinfo {pages} {2626} (\bibinfo {year}
  {2008})}\BibitemShut {NoStop}%
\bibitem [{\citenamefont {Hemmerich}\ \emph {et~al.}(1994)\citenamefont
  {Hemmerich}, \citenamefont {Weidem{\"u}ller},\ and\ \citenamefont
  {H{\"a}nsch}}]{hemmerich94}%
  \BibitemOpen
  \bibfield  {author} {\bibinfo {author} {\bibfnamefont {A.}~\bibnamefont
  {Hemmerich}}, \bibinfo {author} {\bibfnamefont {M.}~\bibnamefont
  {Weidem{\"u}ller}}, \ and\ \bibinfo {author} {\bibfnamefont {T.}~\bibnamefont
  {H{\"a}nsch}},\ }\href@noop {} {\bibfield  {journal} {\bibinfo  {journal}
  {EPL}\ }\textbf {\bibinfo {volume} {27}},\ \bibinfo {pages} {427} (\bibinfo
  {year} {1994})}\BibitemShut {NoStop}%
\end{thebibliography}%

\end{document}